# ELLIPTICAL X-RAY SPOT MEASUREMENT

Roger A. Richardson, Stephen Sampayan, John Weir, LLNL, Livermore, CA, USA

*Abstract*

The so-called roll bar measurement uses a heavy metal material, optically thick to x-rays, to form a shadow of the x-ray origination spot. This spot is where an energetic electron beam interacts with a high Z target. The material (the "roll bar") is slightly curved to avoid alignment problems. The roll bar is constructed and positioned so that the x-rays are shadowed in the horizontal and vertical directions, so information is obtained in two dimensions. If a beam profile is assumed (or measured by other means), the equivalent x-ray spot size can be calculated from the x-ray shadow cast by the roll bar. Thus the ellipticity of the beam can be calculated, assuming the ellipse of the x-ray spot is aligned with the roll bar. The data is recorded using a scintillator and gated camera. Data will be presented from measurements using the ETA II induction LINAC. The accuracy of the measurement is checked using small elliptical targets.

## 1 INTRODUCTION

This paper describes a "rollbar" diagnostic used to determine the spot size of a pulsed x-ray source. The innovation described here is a rollbar in two dimensions, so that the beam ellipticity (with respect to the rollbar axis) and center can be measured. A gated camera is used so that a time history of the spot size can be generated using multiple shots or cameras.

## 2 EXPERIMENTAL SETUP

The diagnostic is basically the same as described in reference 1. A thick block of Tungsten alloy metal is used to image a x-ray spot generated by the focused beam of the ETA II LINAC [2]. The edge of the block has a large (1 meter) radius ("rolled") so that alignment is not critical. The modification described in this paper is that two edges of the rollbar (perpendicular to the line of sight) are radiused, so that the spot size can be determined in two dimensions and the horizontal & vertical components of the beam can be measured. A fit to the data gives the spot size and position of the x-ray spot in these directions.

The shadow of the x-ray spot is projected onto a scintillator (1 cm Bicron BC-400), which is then imaged using an intensified gated camera. The cameras used to take the data were a Cohu SIT tube, with a typically 10ns gate, and/or the DiCAM Pro camera from Cooke, with a 5 ns gate.

The advantage of a rollbar measurement over a true 2D measurement (such as a pinhole) is that the signal can be averaged in one dimension, and photon statistics are vastly improved. These image lineouts are typically fitted to an integrated gaussian function (erfc) to obtain the spot size. The spot size is defined as the full width at half maximum (FWHM) of the gaussian. Alternatively the lineouts can be fitted to a double gaussian, or the FFT can be done.

To measure the resolution of the diagnostic, the rollbar was placed adjacent to the scintillator, which gave a measured edge blur (using an erfc fit) of 2.7 mm. At the radiographic magnification of 4 used to take this data, this gives a resolution of 0.67 mm. The magnification is the ratio of the distances from the rollbar to the scintillator and from the rollbar to the spot.

A sequence of four accelerator shots with identical settings were used to estimate a 7% error in the measured RMS (root mean square) spot size, and a 14% error in the ratio of horizontal/vertical (ellipticity). This measurement is the basis for the error bars on the plots in this paper.

## 3 DATA

### 3.1 Elliptical targets

The accelerator beam is typically focused on a flat, 5 mil tantalum target to produce x-rays detected by this diagnostic. To check the accuracy of this measurement, a target was constructed that consisted of small tantalum buttons, which were glued onto a thin sheet of Mylar. The targets were small (on the order of the spot size of the beam), elliptical in shape, 1 x 1.5 mm, and oriented to align with the rollbar axis ("horizontal" and "vertical"). A successful shot is shown in Fig. 1, which is a 10 ns gated image, centered temporally in the middle of the pulse. The data is somewhat noisy because the outer part of the beam missed the target and created noise x-rays. Gaussian fits of the image lineouts in the horizontal and vertical directions gives a measured spot size of 1.1 x 1.6 mm, which compares well with the target size and shape, especially considering the minimum resolution of ~0.7 mm

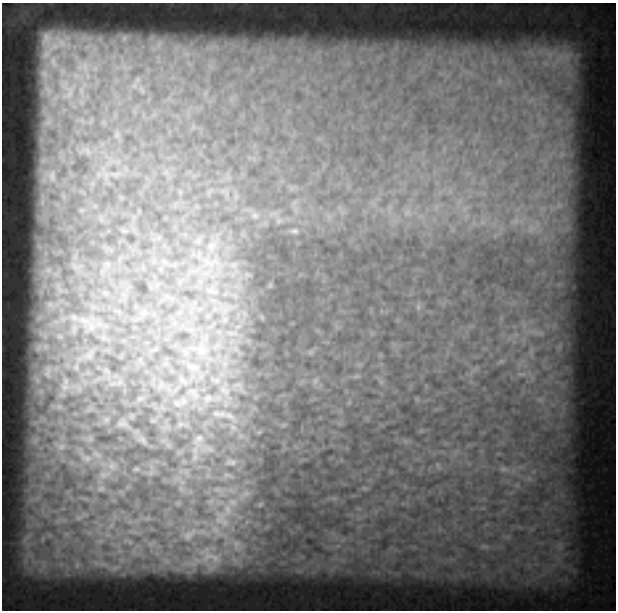

Figure 1: Two-dimensional rollbar image.

### 3.2 Quadrapole magnet variation

A quadrapole magnet in the beamline of the accelerator was varied to adjust the beam ellipticity. This produced an elliptical x-ray spot size as illustrated in Fig. 2.

The electron beam leaves a witness hole in the tantalum disk after a shot. These holes were elliptical in shape during this data series, and the measured ellipticity of the holes linearly increased with the quadrapole current, Fig. 3. The hole shapes also confirmed the direction of the long axis of the ellipse (horizontal). In general, the witness holes do not give an accurate measurement of spot size, however in this case they did correlate well with the ellipticity.

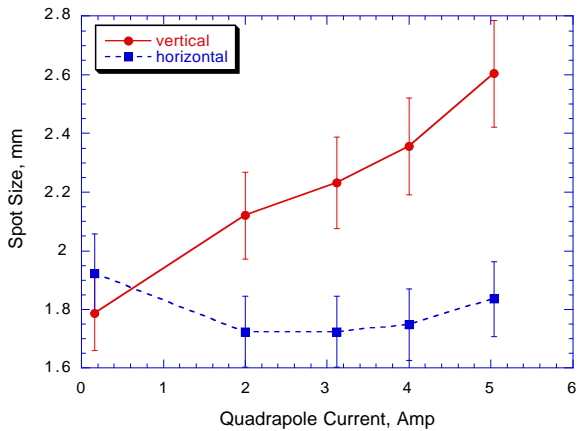

Figure 2: Spot size variation due to quadrapole magnet current.

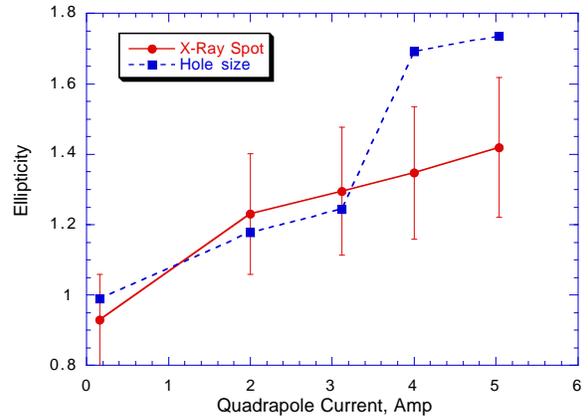

Figure 3: Spot size ellipticity compared to target hole shape.

### 3.3 Tuning curve

The final focus of the accelerator is minimized by adjusting the final magnet and plotting the spot size as determined by the rollbar. A typical tuning curve is shown in Fig. 4. In this tune, the beam is almost round at the spot size minimum, and become elliptical as the spot size increases.

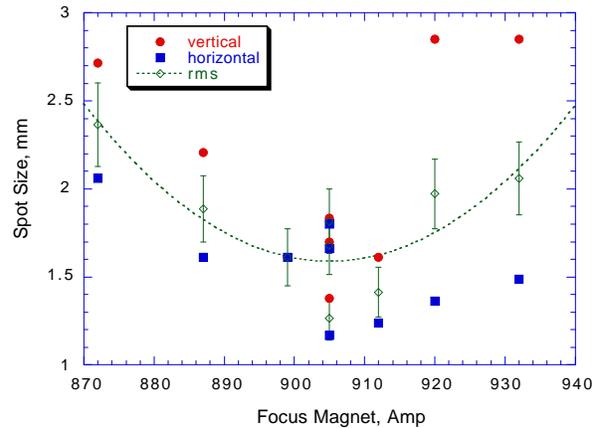

Figure 4: Accelerator tuning curve.

### 3.4 X-Ray spot vs. time

By changing the delay of the gated camera, a time history of the spot size can be constructed with multiple shots. A typical data series is shown in figure 4. This illustrates the beam degradation during the rise and fall of the pulse due to energy variations in the beam.

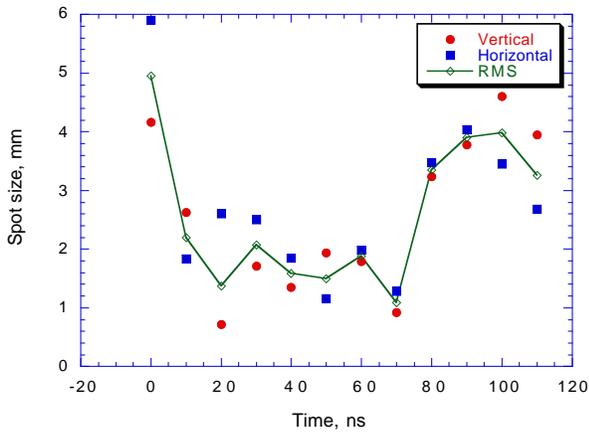

Figure 5: X-Ray spot size vs. time.

*3.5 Beam motion*

The x-ray spot center can also be calculated from the data and plotted. Figure 5 shows some data taken using a gas cell for the final focus [3]. This experiment used a small cell filled with nitrogen gas to focus the beam without using a magnet. The electron beam ionizes the gas in the cell, and the resulting plasma neutralizes the space charge of the beam, causing the beam to pinch due to self-magnetic forces. Because the focusing magnet is turned off, the beam now has freedom of motion on the target and moves during the shot.

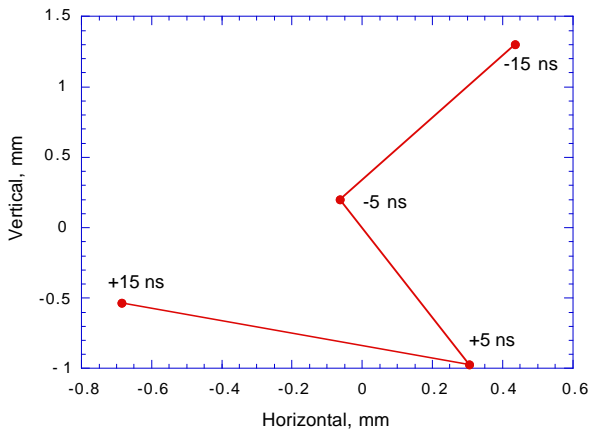

Figure 6: Spot centroid motion. Data points are 10 ns apart, labeled with respect to middle of pulse.

## 4 SUMMARY

The two dimensional rollbar technique gives an accurate measurement of the spot size in two dimensions (in the rollbar orientation). The data can be analyzed to give the x-ray spot ellipticity and centroid. It is a valuable diagnostic for tuning a radiographic accelerator and studying beam-target interactions.

## REFERENCES

[1] Richardson, R. A., "Optical Diagnostics on ETA II For X-ray Spot Size", Proceedings of the PAC'99, New York, NY, (1999).
[2] S.Sampayan, et. al. "Beam-Target Interaction Experiments For Bremsstrahlung Converters Applications", Proceedings of the LINAC2000, Monterey, CA, (2000).
[3] To be published.

## ACKNOWLEDGMENTS

This work was performed under the auspices of the U.S. Department of Energy by University of California Lawrence Livermore National Laboratory under contract No. W-7405-Eng-48.